\documentclass[preprint2]{aastex62}
\usepackage{multirow}
\usepackage{color}

\usepackage{amssymb,amsmath}
\usepackage{natbib}  % Use the "Natbib" style for the
\usepackage{subfigure}  % Use the "Natbib" style for the
\usepackage{rotating}
\usepackage{longtable}
\usepackage[mediumspace,Gray,squaren]{SIunits} % (roman \mu is \micro)
\usepackage{rotating}
\graphicspath{{.}{./figures/}}
  \DeclareGraphicsExtensions{.pdf, .eps}
  \usepackage{epstopdf}
\newcommand{\beq}{\begin{equation}}
\newcommand{\eeq}{\end{equation}}
\newcommand{\be}{\begin{equation}}
\newcommand{\ee}{\end{equation}}
\newcommand{\bea}{\begin{eqnarray}}
\newcommand{\eea}{\end{eqnarray}}
\newcommand{\bdi}{\begin{displaymath}}
\newcommand{\edi}{\end{displaymath}}

\newcommand{\sqdeg}{\ensuremath{{\rm deg}^2}}

\newcommand{\rmicron}{$\,\micro$m}

\newcommand{\HI}{H\,{\sc i}}

\newcommand\lsim{\,\lower2truept\hbox{${<\atop\hbox{\raise4truept\hbox{$\sim$}}}$}\,}
\newcommand\gsim{\,\lower2truept\hbox{${>\atop\hbox{\raise4truept\hbox{$\sim$}}}$}\,}

\newcommand{\Rmnum}[1]{\expandafter\@slowromancap\romannumeral #1@}
\newcommand{\herschelspire}{{\it Herschel}/SPIRE}
\newcommand{\planck}{{\it Planck}}
\newcommand{\msun}{\ensuremath{M_\odot}}
\newcommand{\lsun}{\ensuremath{L_\odot}}
\newcommand{\fieldname}{\textsc{ra23h30dec-55}}

\newcommand{\plkfacA}{0.98} % 350um
\newcommand{\plkfacB}{1.32} % 550um
\newcommand{\plkfacC}{1.64} % 1380um
\newcommand{\plkfacD}{0.423} % 2100um

\defcitealias{george15}{G15}

\shorttitle{Cross Spectra from 95 to 1200 GHz}
\shortauthors{Viero et al.}
\newcommand{\StanfordKPAC}{Kavli Institute for Particle Astrophysics and Cosmology \& Physics Department, Stanford University, Stanford, CA 94305, USA}
\newcommand{\KICPChicago}{Kavli Institute for Cosmological Physics, University of Chicago, 5640 S. Ellis Ave., Chicago, IL 60637, USA}
\newcommand{\AAUChicago}{Department of Astronomy and Astrophysics, University of Chicago, 5640 S. Ellis Ave., Chicago, IL 60637, USA}
\newcommand{\Davis}{Department of Physics, University of California, Davis, CA 95616, USA}
\newcommand{\FNAL}{Fermi National Accelerator Laboratory, MS209, P.O. Box 500, Batavia, IL 60510, USA}
\newcommand{\ArgonneHEP}{High Energy Physics Division, Argonne National Laboratory, Argonne, IL 60439, USA }
\newcommand{\PhysicsUChicago}{Department of Physics, University of Chicago, Chicago, IL 60637, USA }
\newcommand{\EFIChicago}{Enrico Fermi Institute, University of Chicago, Chicago, IL 60637, USA }
\newcommand{\SLAC}{SLAC National Accelerator Laboratory, 2575 Sand Hill Road, Menlo Park, CA 94025, USA}
\newcommand{\McGill}{Department of Physics and McGill Space Institute, McGill University, Montreal, Quebec H3A 2T8, Canada}
\newcommand{\Caltech}{California Institute of Technology, Pasadena, CA 91125, USA}
\newcommand{\Berkeley}{Department of Physics, University of California, Berkeley, CA 94720, USA }
\newcommand{\Cifar}{Canadian Institute for Advanced Research, CIFAR Program in Cosmology and Gravity, Toronto, ON, M5G 1Z8, Canada}
\newcommand{\Colorado}{Center for Astrophysics and Space Astronomy, Department of Astrophysical and Planetary Sciences, University of Colorado, Boulder, CO 80309, USA }
\newcommand{\ESO}{European Southern Observatory, Karl-Schwarzschild-Stra{\ss}e 2, 85748 Garching, Germany}
\newcommand{\Colphys}{Department of Physics, University of Colorado, Boulder, CO 80309, USA}
\newcommand{\Illast}{Astronomy Department, University of Illinois at Urbana-Champaign, 1002 W. Green Street, Urbana, IL 61801, USA}
\newcommand{\Illphys}{Department of Physics, University of Illinois Urbana-Champaign, 1110 W. Green Street, Urbana, IL 61801, USA}
\newcommand{\UChicago}{University of Chicago, Chicago, IL 60637, USA}
\newcommand{\KIPAC}{Kavli Institute for Particle Astrophysics and Cosmology, Stanford University, 452 Lomita Mall, Stanford, CA 94305, USA}
\newcommand{\LBNL}{Physics Division, Lawrence Berkeley National Laboratory, Berkeley, CA 94720, USA }
\newcommand{\Arizona}{Steward Observatory, University of Arizona, 933 North Cherry Avenue, Tucson, AZ 85721, USA}
\newcommand{\Michigan}{Department of Physics, University of Michigan, Ann  Arbor, MI 48109, USA}
\newcommand{\Munich}{Faculty of Physics, Ludwig-Maximilians-Universit\"{a}t, 81679 M\"{u}nchen, Germany}
\newcommand{\ExcellenceCluster}{Excellence Cluster Universe, 85748 Garching, Germany}
\newcommand{\MPE}{Max-Planck-Institut f\"{u}r extraterrestrische Physik, 85748 Garching, Germany}
\newcommand{\Dunlap}{Dunlap Institute for Astronomy \& Astrophysics, University of Toronto, 50 St George St, Toronto, ON, M5S 3H4, Canada}
\newcommand{\Minnesota}{Department of Physics, University of Minnesota, Minneapolis, MN 55455, USA }
\newcommand{\Melbourne}{School of Physics, University of Melbourne, Parkville, VIC 3010, Australia}
\newcommand{\CaseWestern}{Physics Department, Case Western Reserve University,Cleveland, OH 44106, USA }
\newcommand{\ArtInstChicago}{Liberal Arts Department, School of the Art Institute of Chicago, Chicago, IL 60603, USA }
\newcommand{\JPL}{Jet Propulsion Laboratory, California Institute of Technology, Pasadena, CA 91109, USA}
\newcommand{\CfA}{Harvard-Smithsonian Center for Astrophysics, Cambridge, MA 02138, USA }
\newcommand{\Stanford}{Deptartment of Physics, Stanford University, 382 Via Pueblo Mall, Stanford, CA 94305, USA}
\newcommand{\UToronto}{Department of Astronomy \& Astrophysics, University of Toronto, 50 St George St, Toronto, ON, M5S 3H4, Canada}

\newcommand{\Rochester}{Rochester Institute of Technology, Rochester, NY 14623, USA}
\newcommand{\IAS}{Institut d'Astrophysique Spatiale (IAS), B\^atiment 121, F- 91405 Orsay (France); Universit\'e Paris-Sud 11 and CNRS (UMR 8617)}

\begin{document}
%Measuring Cross-Spectra of the Cosmic Infrared Background and Cosmic Microwave Background from 95 to 1200 GHz
\title{Measurements of the Cross Spectra of the Cosmic Infrared and Microwave Backgrounds from 95 to 1200\,GHz}

\correspondingauthor{C.~L.~Reichardt}
\email{marco.viero@stanford.edu, christian.reichardt@unimelb.edu.au}

\author{M.~P.~Viero}
\affiliation{\StanfordKPAC}

\affiliation{\Caltech}

\author{C.L.~Reichardt}
\affiliation{\Melbourne}

\author{B.~A.~Benson}
\affiliation{\FNAL}
\affiliation{\KICPChicago}
\affiliation{\AAUChicago}

\author{L.~E.~Bleem}
\affiliation{\ArgonneHEP}
\affiliation{\KICPChicago}

\author{ J.~Bock}
\affiliation{\Caltech}
\affiliation{\JPL}

\author{J.~E.~Carlstrom}
\affiliation{\KICPChicago}
\affiliation{\PhysicsUChicago}
\affiliation{\ArgonneHEP}
\affiliation{\AAUChicago}
\affiliation{\EFIChicago}

\author{C.~L.~Chang}
\affiliation{\ArgonneHEP}
\affiliation{\KICPChicago}
\affiliation{\AAUChicago}

\author{H-M.~Cho}
\affiliation{\SLAC}

\author{T.~M.~Crawford}
\affiliation{\KICPChicago}
\affiliation{\AAUChicago}

\author{A.~T.~Crites}
\affiliation{\KICPChicago}
\affiliation{\AAUChicago}
\affiliation{\Caltech}

\author{T.~de~Haan}
\affiliation{\McGill}
\affiliation{\Berkeley}

\author{M.~A.~Dobbs}
\affiliation{\McGill}
\affiliation{\Cifar}

\author{W.~B.~Everett}
\affiliation{\Colorado}

\author{E.~M.~George}
\affiliation{\Berkeley}
\affiliation{\ESO}

\author{N.~W.~Halverson}
\affiliation{\Colorado}
\affiliation{\Colphys}

\author{N.~L.~Harrington}
\affiliation{\Berkeley}

\author{G. Holder}
\affiliation{\Illast}
\affiliation{\Illphys}
\affiliation{\McGill}

\author{W.~L.~Holzapfel}
\affiliation{\Berkeley}

\author{Z.~Hou}
\affiliation{\KICPChicago}
\affiliation{\AAUChicago}

\author{J.~D.~Hrubes}
\affiliation{\UChicago}

\author{L.~Knox}
\affiliation{\Davis}

\author{A.~T.~Lee}
\affiliation{\Berkeley}
\affiliation{\LBNL}

\author{D.~Luong-Van}
\affiliation{\UChicago}

\author{D.~P.~Marrone}
\affiliation{\Arizona}

\author{J.~J.~McMahon}
\affiliation{\Michigan}

\author{S.~S.~Meyer}
\affiliation{\KICPChicago}
\affiliation{\AAUChicago}
\affiliation{\EFIChicago}
\affiliation{\PhysicsUChicago}

\author{M.~Millea}
\affiliation{\Davis}

\author{L.~M.~Mocanu}
\affiliation{\KICPChicago}
\affiliation{\AAUChicago}

\author{J.~J.~Mohr}
\affiliation{\Munich}
\affiliation{\ExcellenceCluster}
\affiliation{\MPE}

\author{L.~Moncelsi}
\affiliation{\Caltech}

\author{S.~Padin}
\affiliation{\KICPChicago}
\affiliation{\AAUChicago}

\author{C.~Pryke}
\affiliation{\Minnesota}

\author{J.~E.~Ruhl}
\affiliation{\CaseWestern}

\author{K.~K.~Schaffer}
\affiliation{\KICPChicago}
\affiliation{\EFIChicago}
\affiliation{\ArtInstChicago}

\author{P.~Serra}
\affiliation{\JPL}
\affiliation{\Caltech}
\affiliation{\IAS}

\author{E.~Shirokoff}
\affiliation{\Berkeley}
\affiliation{\KICPChicago}
\affiliation{\AAUChicago}

\author{Z.~Staniszewski}
\affiliation{\CaseWestern}
\affiliation{\JPL}

\author{A.~A.~Stark}
\affiliation{\CfA}

\author{K.~T.~Story}
\affiliation{\KICPChicago}
\affiliation{\PhysicsUChicago}
\affiliation{\KIPAC}
\affiliation{\Stanford}

\author{K.~Vanderlinde}
\affiliation{\Dunlap}
\affiliation{\UToronto}

\author{J.~D.~Vieira}
\affiliation{\Illast}
\affiliation{\Illphys}

\author{R.~Williamson}
\affiliation{\KICPChicago}
\affiliation{\AAUChicago}

 \author{M.~Zemcov}
   \affiliation{\Rochester}
 \affiliation{\JPL}

\begin{abstract}
We present measurements of the power spectra of cosmic infrared background (CIB) and cosmic microwave background (CMB) fluctuations in six frequency bands. 
Maps at the lower three frequency bands, 95, 150, and 220\,GHz (3330, 2000, 1360\,\rmicron) are from the South Pole Telescope, while the upper three frequency bands, 600, 857, and 1200\,GHz (500, 350, 250\,\rmicron)  are observed with \herschelspire. 
From these data, we produce 21 angular power spectra (six auto- and fifteen cross-frequency) spanning the multipole range $600 \le \ell \le $ 11{,}000.   
Our measurements are the first to cross-correlate measurements near the peak of the CIB spectrum with maps at 95\,GHz, 
complementing and extending the measurements from \citet{planck13-30} at 143 - 857\,GHz.  
The observed fluctuations originate largely from clustered, infrared-emitting,  
dusty star-forming galaxies, the CMB, and to a lesser extent radio galaxies, active galactic nuclei, and the Sunyaev-Zel\rq dovich effect.    
\end{abstract}


% ======================
\section{Introduction}
\label{sec:intro}
% ======================
\setcounter{footnote}{0}

Maps of the cosmic microwave background (CMB) provide a snapshot of the early Universe, and lead to some of the most powerful tests of cosmology today \citep[e.g.,][]{hou18, henning18, planck18-6, louis17}. 
The CMB is by far the brightest signal at millimeter (mm) wavelengths from scales of a few degrees to few arcminutes. 
However, on smaller angular scales or at higher frequencies, the maps begin to be dominated by a combination of the thermal and kinematic Sunyaev-Zel'dovich (SZ) effects, radio galaxies, and thermal emission from dusty star-forming galaxies \citep[DSFGs;][]{hauser01, lagache05,casey14}. 
This complex mixture encodes information about structure growth and the connection between galaxies, star formation, and the underlying dark matter haloes. 

DSFGs are actively star-forming galaxies (with typical star formation rates $>10\,{\rm \msun/yr}$ and $\rm L_{IR} > 10^{11} \,\lsun$), enshrouded at varying levels by interstellar dust grains \citep{draine84, draine03}.   
 The dust efficiently absorbs the ultraviolet light from the hot, young stars and thermally re-radiates as a modified blackbody with a characteristic temperature of $\rm T_d\sim30\,K$ and a peak in emission at $\rm \lambda_{rest} \sim 100\,\mu m$.
Thermal emission from the DSFGs is collectively referred to as the cosmic infrared background (CIB). 
The CIB accounts for approximately half of the total energy released over the history of cosmic star formation \citep{dole06}.   

Despite the brightness of the CIB and the correspondingly abundant population of DSFGs, directly resolving a typical DSFG   
has proven extremely challenging, 
mostly because of  
the blending together of sources by 
large instrumental beams 
\citep[source confusion; e.g.,][]{blain98a,lagache03,nguyen10}.  
While the most luminous sources are often easily identified \citep[e.g.,][]{smail97,barger98,hughes98c} --- even at very high redshifts \citep[e.g.,][]{marrone18} --- these luminous DSFGs make up at most 15\% of the CIB \citep{viero13b}.  
Most of the remaining intensity is found in  
\lq\lq luminous infrared galaxies\rq\rq, a class of galaxies that are become increasingly common at $z\gsim 0.5$, with infrared luminosities of log$(L/\rm L_{\odot})=11-12$, star-formation rates of $10-100\, \rm M_{\odot}\, yr^{-1}$, stellar masses of log$(M/\rm M_{\odot})=10-11$ \citep[generally around the knee of the stellar-mass function, M$_{\star}$; e.g.,][]{ilbert13,muzzin13a}, 
and 600\,GHz flux densities of $1-5\, \rm mJy$.  
The only way to measure the signal from sources this faint is statistically. Generally this is done in real space through the covariance of an image with ancillary data \citep[also known as stacking; e.g.,][]{dole06,marsden09,viero12,viero13b}; or in Fourier space through correlations in the spatial distribution of their projected emission \citep[e.g.,][]{bond86,haiman00,knox01}.

%WHAT HAS COME BEFORE.   
The first detection of galaxy clustering in the CIB anisotropy was made at 1.9\,THz (160\rmicron) in \emph{Spitzer} data \citep{grossan07, lagache07}, followed by measurements at 600, 857, and 1200\,GHz with BLAST \citep{viero09}, and at 150 and 220\,GHz with SPT \citep[][]{hall10}.  
Subsequent observations with \emph{Herschel}/SPIRE at 600, 857, and 1200\,GHz \citep{amblard11, viero13a, thacker13}  
confirmed and extended these measurements to include cross-frequency spectra. %, while at 
Other works began connecting the millimeter and submillimeter backgrounds, using data from  \emph{Planck} at 217, 353, 545, and 857\,GHz \citep[][]{planck11-18}, or from a combination of BLAST and the Atacama Cosmology Telescope (ACT) at 148 and 218\,GHz  \citep[][]{hajian12}. 
At millimeter wavelengths, ACT \citep{dunkley11} and SPT \citep[][ hereafter G15]{reichardt12b,george15} robustly measured power originating from DSFGs and radio galaxies, with  \citetalias{george15} placing constraints on the level of correlation between CIB and thermal SZ (tSZ) of $\xi = 0.113\pm 0.056$.

%WHAT WE WANT TO DO AND WHY?   
In this work we build upon previous efforts bridging the CIB and CMB by combining 
millimeter data observed with the South Pole Telescope \citep[SPT;][]{carlstrom11} at 95, 150, and 220\,GHz, 
and submillimeter data imaged with the SPIRE instrument \citep{griffin10} aboard the \emph{Herschel Space Observatory} \citep{pilbratt10} at 600, 857, and 1200\,GHz.   
Our measurements complement those made on larger scales by \emph{Planck} --- which are particularly sensitive to the linear clustering term --- by adding data on smaller angular scales, where non-linear clustering and shot noise in the galaxy counts dominate.  
We find strong correlations in and between maps at all wavelengths, increasing the number of bands where correlated signal is robustly detected.

%OUTLINE
The paper is organized as follows. 
In Section~\ref{sec:data} we describe the SPIRE and SPT data and lay out how we estimate bandpowers.   
In Section~\ref{sec:bandpowers} we present the bandpowers and quantify the degree of correlations between observing frequency bands.     
We summarize our results  in Section~\ref{sec:conclusion}. 

% ======================
%\section{Maps}
\section{Data and Analysis}
\label{sec:data}

%short summary
We present power spectrum measurements based on observations of roughly $100\, \rm deg^2$ of sky in six frequency bands with \emph{Herschel}/SPIRE and the SPT. 
Specifically, 102\,\sqdeg{} are used for the SPT-only bandpowers, 86\,\sqdeg{} are used for the $\rm SPT\times SPIRE$ bandpowers, and 90\,\sqdeg{} are used for the SPIRE-only bandpowers. 
Bandpowers are estimated using a pseudo-$C_{\ell}$ 
method, with the covariance matrix 
estimated with a combination of the scatter in the data and Monte Carlo simulations.

The SPT was used to observe the field at 95, 150 and 220\,GHz; maps and power spectra from the SPT frequency bands will hereafter be referred to as SPT. 
The \emph{Herschel}/SPIRE maps are at 600, 857, and 1200\,GHz (500, 350, and 250\,\rmicron), and will be labeled by SPIRE. 
We will refer to cross-spectra between the sets as $\rm SPIRE\times SPT$.

%#######################
\subsection{SPT Data and Power Spectra}
\label{sec:spt}
% #######################
Between 2008 and 2011 the SPT-SZ camera on the SPT was used primarily to survey $2540$ deg$^2$ of the sky at 95, 150, and $220\, \rm GHz$. 
The survey region covers declinations between -40$^\circ$ and -65$^\circ$ and right ascensions from $20\, \rm h$ to $7\, \rm h$. 
This work is based on one of two deep fields that have $\sqrt{2}$ lower noise at 150\,GHz than the full survey; deep field noise levels are 
$\lsim 13\, \micro \rm K$-arcmin at 150\, GHz. 
The field used in this work  
is approximately 10$^\circ$x10$^\circ$ centered at $\rm 23^h30^m, -55^d00^m$, and was observed in 2008 and 2010. 
A full list of the SPT-SZ fields can be found in \citet{story13}. 
   
   The SPT-SZ data,  maps, and power spectrum analysis used in this work are described by \citetalias{george15}. 
   Briefly, the data are band-pass filtered to remove atmospheric emission at low frequencies and to avoid aliasing high-frequency noise into the signal band. 
   The data are then binned into 0.5$^\prime$ pixels, and the power spectrum estimated using a pseudo-$C_\ell$ method \citep{hivon02, polenta05, tristram05}.
   We emphasize that except for the calibration factor the CMB-only bandpowers in this work are identical to the bandpowers for this field in \citetalias{george15}, and should therefore only be combined with the publicly available \citetalias{george15} bandpowers that do not include the \fieldname{} field. 
   
   %Beams and calibration
The SPT beams are determined as in \citetalias{george15}. 
Briefly, the beam is measured using observations of Venus and Jupiter (see, e.g., \citealt{story13}). 
The approximate FWHMs of the beam mainlobes are  $1.\rq 7$, $1.\rq 2$, and $1.\rq 0$ at 95, 150, and $220\, \rm GHz$, respectively. 
The absolute calibration of the maps has been updated since \citetalias{george15}. 
As described by  \citet{hou18}, the new calibration is determined by comparing the SPT maps to \planck{} maps over the same footprint. 
%From TC: 2018/9/15
The uncertainty in the SPT power calibration at [95, 150,  220] GHz is [0.43\%, 0.34\%, 0.84\%].  
The correlated part of these uncertainties is 0.34\%. 
The beam and calibration uncertainties are included as a correlation matrix in the data release.

%DESCRIBE COADDS HERE!!!!
For estimating the $\rm SPT \times  SPIRE$ bandpowers, we also coadd the $O$(100) complete, independent SPT maps used by \citetalias{george15} into a single map at each SPT frequency. 
Each individual map is assigned equal weight when co-adding to match the weighting used in the $\rm SPT \times  SPT$ bandpowers. 
By coadding, we reduce the number of FFTs required in the bandpower estimation. 
Using a single coadded SPT map does not incur a noise bias as the SPT and SPIRE noise is uncorrelated. 
We will discuss the $\rm SPIRE \times  SPT$ bandpower estimation more in the next section.

%#######################
\subsection{Herschel/\emph {SPIRE} Data}
\label{sec:spire}
% #######################
The SPIRE maps at 600, 857, and 1200\,GHz (500, 350, and 250\,\rmicron) used in this work
are the same as those used in
\citet{holder13} with minor modifcations, described below. The maps were constructed from observations   
designed to target large scales by observing in fast-scan mode (60\, arcsec\, sec$^{-1}$) to minimize the impact of $1/f$ noise.   
The full observation consists of 
twelve $10\times 3.3$ degree scans, resulting in four complete passes of the entire region.  
The scan direction is alternated such that half of the scans are perpendicular to the other half. 
The $100\, \rm deg^2$ map is constructed from a mosaic of these twelve scans.

Data are reduced with standard ESA software and the custom software package, {\sc SMAP} \citep{levenson10, viero13a}.   
Maps are made using 
 SMAP/SHIM \citep{levenson10}, an iterative map-maker designed to separate large-scale noise from signal.   
 Initial processing uses the Herschel Interactive Processing Environment (HIPE\footnote{\url{http://herschel.esac.esa.int/hipe/}}) 
release version 13.0.4887, which includes that calibration tree version spire\_cal\_12\_3.   
Practically, this represents a change in the calibration from  maps used in previous SPT-SPIRE studies \citep[e.g.,][]{holder13,hanson13} of 1.1, 1.3, and 1.3\% at 600, 857, and 1200\,GHz, respectively.

Two sets of maps with different pixel sizes 
are made: one with pixels 15\arcsec\ on a side, and the other with pixels 30\arcsec\ on a side. 
The smaller pixels are used for the $\rm SPIRE\times SPIRE$ analysis, in which the 
target signal is detected strongly at scales smaller than 30\arcsec; this is not the case
for $\rm SPIRE\times SPT$ (because of the larger SPT beams), so we can save computing time and storage by using coarser
maps for $\rm SPIRE\times SPT$.

The time-ordered data are split equally into two halves and made into two separate maps, each of which covers the full area.   Hereafter,  these half-data maps are referred to as \emph{jackknife} maps or map pairs, and full-data maps as \emph{co-added}.   
Note, one of the 12 scans is discarded because of a voltage bias swell \citep[or \lq\lq cooler-burp\rq\rq ][]{sauvage14} that occurred during a cooler cycle, which if included would lead to large-scale striping. 
The result is that one of the jackknife maps is not fully sampled; this is taken into account in all stages of bandpower and uncertainty estimation.  
All sets of maps are made in Lambert azimuthal equal-area projection (also known as zenithal equal area, ZEA), with astrometry identical to that of the SPT map.  

The SPIRE map noise levels are $5.6, 4.0$, and $2.8$\,mJy-arcmin, and the instrument effective beams are 36.6\arcsec, 25.2\arcsec, and 18.1\arcsec\ FWHM at 600, 857, and 1200 GHz, respectively.   
The maps are  converted from their native units of $\rm Jy\, beam^{-1}$ to $\rm Jy\, sr^{-1}$ by dividing them by the effective beam areas (see below). 
The absolute calibration uncertainty is 5\%, and is accounted for in the reported beam and calibration correlation matrix. 

SPIRE beams are measured with Neptune; with an angular diameter of $\lsim 2.5\arcsec$, Neptune is effectively a point source to SPIRE.  
Because the effective beam is mildly sensitive to the source spectrum through the passband, the effective solid angle of the beam must be color-corrected \citep{bendo13, griffin13}.  
Following the SPIRE Observers Manual\footnote{Table~5.5 of HERSCHEL-DOC-0798, version 2.5; which can be found at \url{http://herschel.esac.esa.int/Docs/SPIRE/html/spire\_om.html}} for a greybody spectrum representative of the CIB  \citep[$T=13.6\, \rm K$ and $\beta = 1.4$; e.g.,][]{gispert00,planck13-30}, and fully illuminated beam (passbands scaled by $\nu^{1.7}$), the effective solid angles are 3.878, 1.873, and $1.080\times 10^8\, \rm sr$, at 600, 857, and 1200\,GHz, respectively.  
Note that these differ slightly from the solid angles in \citet{viero13a}, which were 3.688,  1.730,  $1.053\times 10^{-8}\, \rm sr$.\footnote{Across the multipole range $\ell=1000 - 2000$, the changes to HIPE and beam solid angles amount to multiplying the bandpowers presented by \citet{viero13a} by factors of 1.05, 0.93 and 0.99 at 600, 857 and 1200\,GHz respectively.}
As described in detail in \citet{viero13a},  the beam power spectra are estimated from the Neptune maps after masking background sources with flux densities greater than 30\,mJy at 1200\,GHz and all pixels beyond 10$\times \rm FWHM$ radius.  
We check that the exact flux density of masked sources and choice of masking radius do not significantly bias the measurement, and include the differences as part of the Monte Carlo procedure described in Section~\ref{sec:errors}.

Lastly, the effective band-centers for the SPIRE bands with fully illuminated beams and a CIB spectral energy distribution (SED) are 546.0, 796.0, and 1092.4\,GHz (549.5, 376.9, and 274.6\,\rmicron). 
The SPT bands have effective CIB band-centers of 99.0, 154.2, and 219.6\,GHz (3.030, 1.946, and 1.367\,mm) respectively.

%%%%%%
%#######################
\subsection{Estimating Bandpowers} 
\label{sec:method}
%#######################

%%%%%%%%%%%%%%%%%%%%%%%%%%
%\begin{sidewaysfigure*}[ht!]
\begin{figure*}[ht]
\centering

\includegraphics[width=\textwidth]{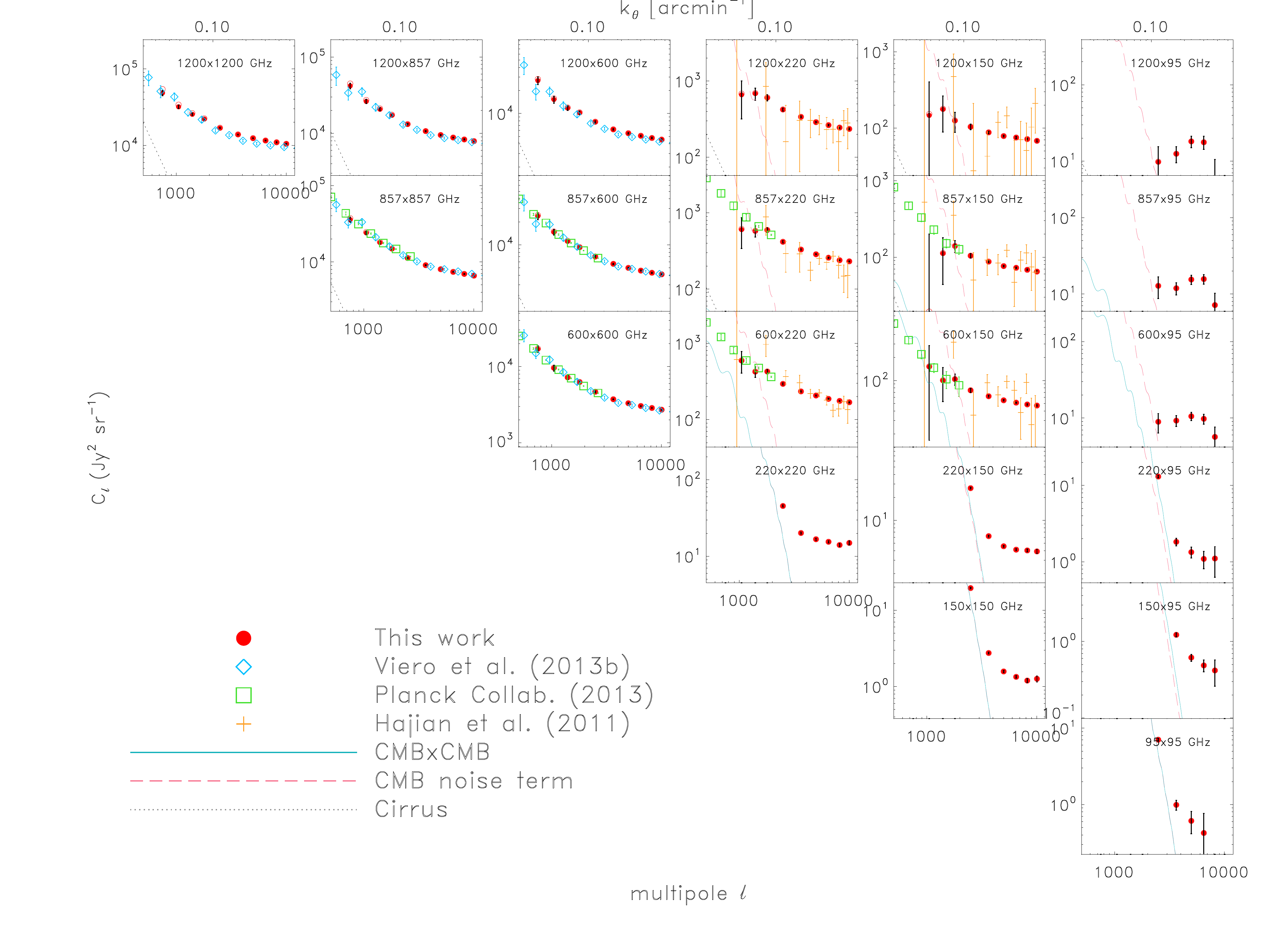} % spt_cibXcib_and_cibXcmb_w_simulated_errors_gaps_20072014.eps}
\caption{Auto- and cross-frequency power spectra of the sky (solid red circles with black error bars), after accounting for mode-coupling, transfer function and instrumental beams. 
Bright point sources have been masked as described in \S\ref{subsec:mask} and an estimate of the Galactic cirrus power subtracted. 
The largely occluded open circles show the results before cirrus-subtraction.   
 Gray dashed lines show the modelled level of Galactic cirrus in each cross-spectrum, while light blue lines show the expected CMB power. 
 The variance due to the CMB (shown by the red, dashed line) can be higher, since it includes a term that scales with the CMB power in each band as well as the cross-spectrum. 
The plotted error bars are derived from the diagonal of the covariance matrix, as described in Section~\ref{sec:errors}; there are significant correlations between bandpowers. 
For this reason and because the spectra change slowly with angular scale, we have rebinned the bandpowers to wider $\ell$-bins for plotting purposes. 
Also shown are measurements from \citet[][blue diamonds]{viero13a}, \citet[][orange crosses]{hajian12}, and the \citet[][green squares]{planck13-30}. 
Note that to compare \emph{Planck} data at 143, 217, 545, and 857\,GHz to that of SPIRE/SPT at 150, 220, 600, and 857\,GHz, the \emph{Planck} values are multiplied by 
\plkfacD, \plkfacC, \plkfacB, and \plkfacA{} respectively. 
}
\label{fig:raw_ps}
\end{figure*}
%%%%%%%%%%%%%%%%%%%%%%%%%%

Bandpowers are estimated using cross-spectrum-based pseudo-$C_{\ell}$ methods \citep{hivon02,tristram05}. 
There are minor implementation differences between the analysis of each of the datasets: SPT-only, SPIRE-only, and $\rm SPT \times SPIRE$. 

The SPT-only and SPIRE-only bandpowers (6 frequency combinations each) are calculated as in previous  SPT and SPIRE power spectrum papers (see \citealt{george15} for the former, and \citealt{viero13a} for the latter). 
To match \citealt{george15}, the SPT-only bandpowers start at $\ell>2000$. 
At larger angular scales, the SPT data are dominated by the CMB. 
The $\rm SPT\times SPIRE$ bandpowers are calculated similarly to the SPIRE only case; 
however, full co-added maps from each experiment are used in this case since the noise in SPT and SPIRE maps are uncorrelated.  
We outline these methods below.

%%%%%%%%%%%%%%%%%%%%%%%%%%
\iffalse
\begin{figure}[htb]
\centering
\includegraphics[width=0.460\textwidth]{spt_cibXcib_and_cibXcmb_Cl_3000_cmb_subtracted_10302014-eps-converted-to.pdf} 
\hspace{3mm}
\caption{$C_{\ell}(\ell=3000)$ levels (symbols) overlaid with their best-fit modified blackbodies (lines).
The total spectra (diamonds and solid line) represents the sum of the Poisson (triangles and dotted line), 2-halo (squares and dashed line), and 1-halo (exes and dot-dashed line) terms.  }
\label{fig:correff}
\end{figure}
\fi

%#######################
\subsubsection{Apodization mask and mode-coupling matrix}
\label{subsec:mask}
%#######################

For each cross-frequency spectrum we construct a unique mask {\bf W} which has values of unity where the two maps overlap, zero outside, and the boundary is tapered with a $1^\circ$ Hanning function. 
Three sets of sources are also masked: 
(1) sources with flux densities greater than 300\,mJy at 1200\,GHz;
(2) local, extended sources (e.g., resolved IRAS sources); and 
(3) sources with flux densities greater than 6.4\,mJy at 150\,GHz.
Sets 1 and 2 are masked from any spectrum involving a SPIRE map, set 3 is masked from any spectrum involving an SPT map.  
As in previous SPT power spectrum measurements  
(e.g., \citealt{reichardt12b,george15}), 
source masks consist of an inner disk that is tapered outside the disk with a Gaussian taper of width $\sigma_{\rm taper} = 5^\prime$. 
The inner disk has radius $5^\prime$ for sources above 50\,mJy at 150\,GHz, $2.5^\prime$ for sources above 6.4\,mJy at 150\,GHz, and zero for sources selected at 1200\,GHz. 
This procedure is followed for each masked pixel in the extended source case.
Finally, the maps are padded with zeros to final sizes greater than twice the original map size.   

To correct for the expected mode-coupling induced by the finite sky coverage,  we analytically calculate 
the mode-coupling matrix $M_{kk^{\prime}} [{\bf W}]$. 
The binned mode-coupling matrix is inverted in the final step of pseudo-$C_{\ell}$ methods to recover an unbiased estimate of the true power spectrum.   

\subsubsection{Transfer function and beams}
\label{sec:tf}

In addition to the mode-coupling induced by the partial sky coverage and removal of bright point sources, we must account for the effects of filtering during map-making and the instrumental beams.  
This filtering could include, for example, high-pass filtering to reduce the impact of $1/f$ noise on the final maps, or polynomial fits to temperature data iteratively removed by HIPE. 
For SPIRE, minimal high-pass filtering is required to make well-behaved maps, as the noise properties are relatively well behaved on large scales  
 \citep[e.g.,][]{pascale11}. 
SPT, on the other hand, suffers from large-scale atmospheric noise, which  
is dealt with by (i) removing a polynomial from each constant-elevation scan and (ii) removing a common mode across each detector wafer for each time sample. 

The transfer functions are determined with the iterative method described in \citet{hivon02} on simulated maps which were made with the same map-making pipelines on mock data.   
They are estimated from 400 simulations of each field and wavelength. 
A unique set of transfer functions is measured for SPIRE, SPT, and $\rm SPIRE\times SPT$. 
We have confirmed that the transfer functions are insensitive to the steepness of the input spectra, and that the transfer functions have converged. 

In addition to the beams, we include the effects of the finite pixel size in the maps according to the approximation in \citet{wu01}: $0.5^\prime$ for maps used in cross-spectra involving SPT data and $0.25^\prime $ for the \emph{Herschel} maps used in the \emph{Herschel} auto-spectra.

%#######################
\subsubsection{ Covariance matrix}
\label{sec:errors}
%#######################

The covariance matrix is estimated in one of two ways depending on the frequency combination. 
In all cases, we assume that the instrumental noise in each frequency band is independent. 
For the $\rm SPT\times SPT$ bandpowers, the covariances are calculated exactly as in \citetalias{george15}; sample variance is estimated from simulations, and  the sum of noise and signal-cross-noise variance is estimated from the scatter in cross-spectra. 
The covariance estimator changes when SPIRE data appears in the covariance element. 
Covariance elements that include SPIRE data are calculated through a Monte-Carlo simulation that creates many realizations of the sky signal and noise terms. 
These noise realizations are drawn from the null-map PSD  for each frequency. 
The sky realization is divided into terms simulated as Gaussian fields (CMB, tSZ, kSZ, radio galaxies) and non-Gaussian fields (CIB). 
The Gaussian terms are fully described by \citetalias{george15}. 
The  CIB mock maps are made with correlated sources such that their clustering signal resembles that of the CIB, with fluxes drawn from the \citet{bethermin12} model.    
Note that the Gaussian and non-Gaussian terms are assumed to be uncorrelated. 
Each sky realization is filtered identically to the real data and added to a noise realization to create mock maps. 
The power spectrum of the mock maps is calculated with the same pipeline as that used for the real data.

The ensemble of estimated binned power spectra, $P_b$, are used to measure, ${\bf V}$, the covariance matrix  
\begin{equation}
{\bf V}_{bb^{\prime}}=\left \langle \left(P_b -  \tilde{P}_b \right)  \left(P_{b^{\prime}} - \tilde{P}_{b^{\prime}}\right) \right \rangle_{{\rm MC}}, 
\end{equation}
where the tilde denotes the mean over all realizations in bin $b$.  
The reported errors in the tables and plots are based on the diagonal entries of ${\bf V}$: 
\begin{equation}
\sigma_{P_{b}}=\sqrt{{\bf V}_{bb}}.
\label{eqn:diagonal}
\end{equation}

The non-Gaussianity of the CIB realizations leads to substantial off-diagonal terms in ${\bf V}$.   
We follow \citet{fowler10}, and check that the simulations are realistic by testing that the four-point function measured directly from data is consistent with that of the simulations.  
We find that the non-Gaussian contribution ranges be between 5 and 10\% of the total error in the regime where the Poisson term dominates.

We also test the robustness of the covariance estimate by splitting the field into four quarters. 
One quarter is not used due to the voltage bias swell mentioned in \S\ref{sec:spire}. 
We estimate bandpowers from null maps created by differencing combinations of the other three quarters, 
and we calculate a $\chi^2$ using the covariance matrix.  
We find a reasonable $\chi^2$ for the number of degrees of freedom (303, 244, 303 respectively for 297 d.o.f) , lending confidence to the covariance estimate.

The reported bandpower error bars and covariance matrix do not include the SPT or SPIRE absolute calibration uncertainties nor the beam uncertainties. 
The released data files do include a correlation matrix encapsulating these uncertainties that should be combined with the model bandpowers and added to the covariance matrix when evaluating the quality of fit to a model.

%#######################
\subsubsection{Handling Galactic Cirrus}
\label{sec:cirrus}
%#######################
The survey region was carefully chosen %specifically 
to minimize Galactic cirrus; nonetheless, it is important to quantify the small cirrus contribution. 
We adopt an approach similar to that taken by \citet{viero13a}.
As described in \S\ref{sec:cirrus_ps}, this approach assumes that the cirrus can be described by a single map (i.e. 100\% correlation between observing bands) with an amplitude that scales with frequency as a modified black-body.  
While a modified black-body spectrum has two free parameters, $T_{\rm eff,c}$ and $\beta_c$, the two are highly degenerate for the current set of observing frequencies. 
Therefore, as we are mostly interested in obtaining a good fit for the purposes of removal (as opposed to the exact qualities of the cirrus) we fix $\beta_c = 1.8$ \citep[e.g.,][]{martin10}.  
Furthermore, we assume that the cirrus spatial power spectrum follows a power law with index $\alpha_{\rm c}$. 
Under these assumptions, the cirrus power in all bands can be fit with two variables, $\alpha_{\rm c}$ and $T_{\rm eff,c}$. 

To determine the power law index, we take advantage of the fact that 
Galactic cirrus is well-traced by nearby 
\HI\ in the Galactic disk \citep[e.g.,][]{boulanger96}, 
with velocities $|v_{\rm LSR}| < 30\, \rm km\, s^{-1}$ with respect to the local standard of reference,   
and column densities $N_{\rm H} \le 5 \times 10^{20}\, \rm cm^{-2}$  
\citep[e.g.,][]{viero14}. 
H$_2$ begins to form at higher column densities $N_{\rm H} \le 5 \times 10^{20}\, \rm cm^{-2}$ \citep{planck13-30}, causing the correlation to break down. 
This field however is of low enough Galactic latitude and column density\footnote{The field's mean column density is $N_{\rm H} = 1.0\times 10^{20}\, \rm cm^{-2}$ and peak is $N_{\rm H}$ of $2.3 \times 10^{20}\, \rm cm^{-2}$.} that this is not an issue.     
We take \HI\  maps from Parkes Galactic All-Sky Survey survey \citep[GASS;][]{mclure09,kalberla10}, and measure the power spectrum of the local velocity component of \HI\ across the $100\, \rm deg^2$ field.
We find that the \HI\ exhibits a power law behavior at $\ell < 600$ (after which the beam attenuates the signal) with slope $\alpha_{\rm c} = -3.1\pm1.4$. 
We assume that that behavior can be extrapolated to larger $\ell$, noting that given the steep fall off there is very little cirrus power in practically all of the reported bandpowers. 
%of approximately $\ell^{-3.1}$ there is very little cirrus power at $\ell>600$. 
We also check that contributions from the higher velocity components of the velocity cube are negligible.

We then perform a Monte Carlo simulation to estimate the cirrus contribution to the SPIRE/SPT bands, and in effect the cirrus correction and its associated uncertainties.  
For each iteration, the cirrus slope and temperatures are drawn from normal distributions of $\alpha_{\rm c} = -3.1 \pm 1.4$ and  $T_{\rm eff,c} = 19.0\pm 1.5\, \rm K$, respectively, and then the spectra are fit with a cirrus power law, templates for the clustered CIB power,\footnote{We use the 1- and 2-halo model templates from \citet{planck11-18}.}
 and a Poisson level.   
In addition to SPIRE and SPT we include spectra at 100\rmicron\ --- where cirrus has a much higher contribution relative to CIB \citep[][]{penin12b} --- from reprocessed IRAS \citep[][]{neugebauer84} observations\footnote{\url{http://www.cita.utoronto.ca/\~mamd/IRIS/IrisDownload.html}} \citep[IRIS;][]{miv05}.  
Note that we assume the IRIS transfer function on these scales is unity \citep[][]{hajian12,penin12b}.   
The variation in the residual cirrus power (after subtracting the mean) is added to the bandpower covariance matrix. 

The cirrus levels are shown as dashed gray lines in Figure~\ref{fig:raw_ps}.   
We find the corrections overall to be very small, 
with the only significant corrections in $\rm SPIRE \times SPIRE$ spectra including 250\rmicron. 
Even in these spectra, the correction is only significant in the lowest-$\ell$ bin and peaks at 
24\% in the first bin of the 250\rmicron{} auto-spectrum.

% ======================
\section{Bandpowers}
\label{sec:bandpowers}
% ======================

The analysis described above is applied to the $\sim100\, \rm deg^2$ of sky in common between the \emph{Herschel}/SPIRE and SPT-SZ surveys. 
 The resulting bandpowers are shown in Figure~\ref{fig:raw_ps} and tabulated for $\rm SPT\times SPT$, $\rm SPIRE\times SPT$, and $\rm SPIRE\times SPIRE$ in Tables~\ref{tab:bandpowers}, \ref{tab:bandpowers2}, and \ref{tab:bandpowers3}, respectively. 
 They are reported in units of Jy$^2$/sr with the photometric convention $\nu I_\nu =$ constant. 
 Point sources  with a 250\rmicron\ flux $\gtrsim 300\, \rm mJy$  are masked in the SPIRE maps, while sources with $150\, \rm GHz$ flux   $\gtrsim 6.4\, \rm mJy$ are masked  in the SPT maps before calculating the bandpowers. 
The bandpowers, covariance matrix, beam and calibration correlation matrix, and window functions are available for download on the SPT\footnote{http://pole.uchicago.edu/public/data/viero18/}.

Cirrus-subtracted power spectra are shown as red circles with black error bars in Figure~\ref{fig:raw_ps}.  
A Poissonian distribution of galaxies at random positions would yield constant power as a function of $\ell$ in this plot. 
The upturn in power towards lower $\ell$ (larger angular scales) is due primarily to the clustering of  DSFGs, with the expected $\rm tSZ\times CIB$ anti-correlation also modifying the observed signal in the $\rm SPIRE\times SPT$ bands. 

At the SPT frequencies, these bandpowers include power from the CMB, tSZ and kSZ, radio galaxies and CIB. The situation simplifies towards higher frequencies where the CIB becomes dominant, however properly modelling the DSFGs of the CIB touches upon a number of open questions about the history of star formation in the Universe and what kind of dark matter haloes host star-forming galaxies \citep{planck13-30, mak17}. 
Undertaking this modelling effort is beyond the scope of this work, however we review some of the basics in Appendix \ref{app:signals}.

%#######################
\subsection{Correlation Coefficients}
\label{sec:correlationcoefficients}
%#######################

%%%%%%%%%%%%%%%%%%%%%%%%%%
\begin{figure*}[ht]
\centering
%\vspace{-90mm}
%\hspace{-8mm}
\includegraphics[width=\textwidth]{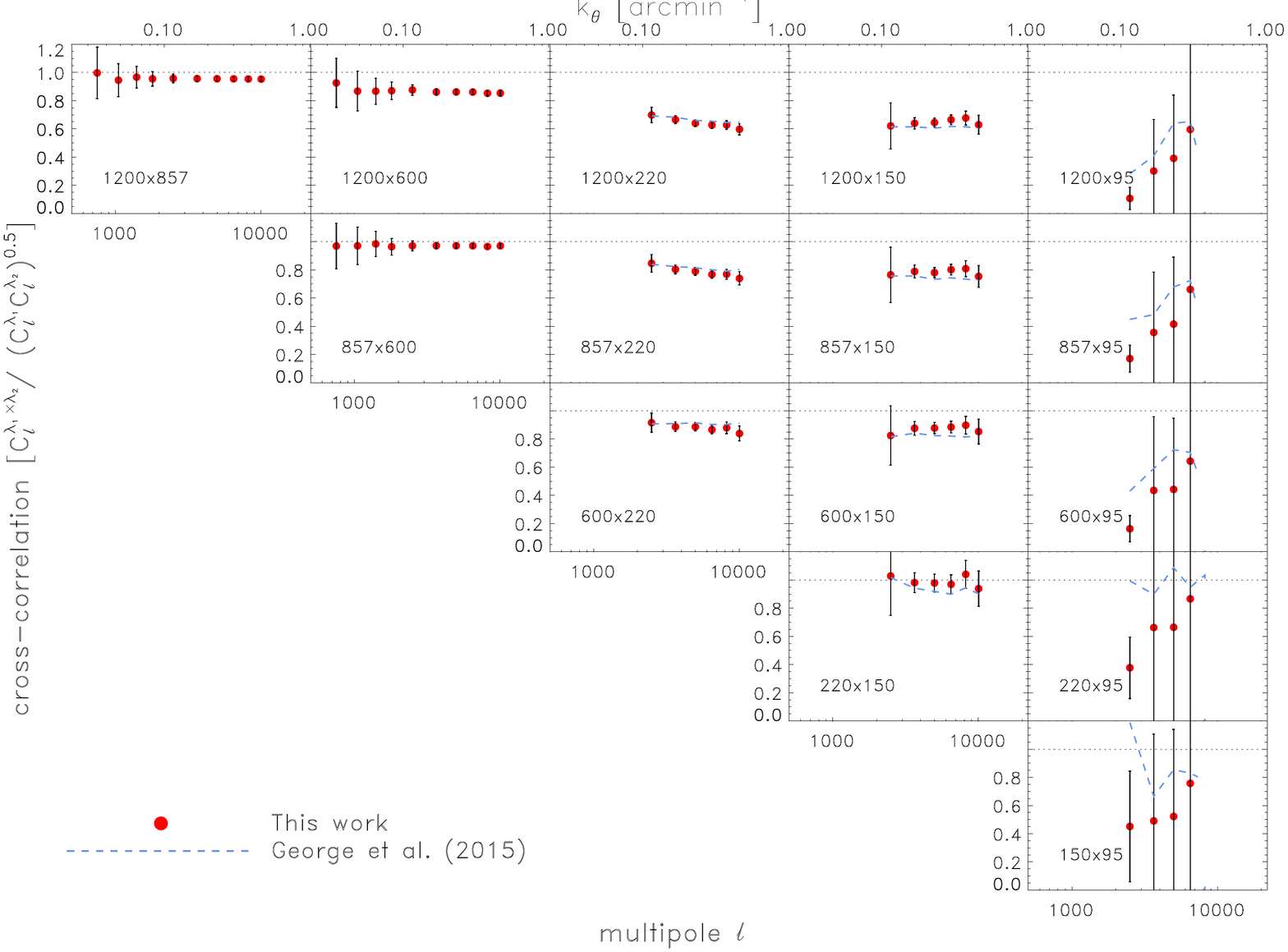}
\caption{Correlation levels between bands for the ``CIB-only'' spectra (see \S\ref{sec:correlationcoefficients} for details).  
Dotted lines represent unity, or what would be expected for two identical maps.  
For spectra including SPT bands, we also show (with a dashed blue line) the correlation predicted by the best-fit model in the baseline analysis of \citetalias{george15}. 
The agreement is excellent at the SPT frequencies with the exception of the largest angular scales at 95\,GHz. 
Recall that this work uses only $\sim$\,100\,\sqdeg{} of the 2500\,\sqdeg{} analyzed by \citetalias{george15}.    
The agreement is also poor for the SPIRE-only frequencies since the simplified CIB model in \citetalias{george15}  has no variation between galaxy SEDs across the DSFG population.  
This simplified model clearly breaks down in the high S/N measurement of CIB power from SPIRE. 
}
\label{fig:xcorr}
\end{figure*}
%%%%%%%%%%%%%%%%%%%%%%%%%%

One question that may be addressed with these data is the degree to which the CIB is correlated across frequencies. 
We define a correlation coefficient between  maps at different frequencies to be 
\beq\label{eqn:correl}
\xi_{\ell}= \frac{C_{\ell}^{\nu_1 \times \nu_2, \prime}}{\sqrt{C_{\ell}^{\nu_1,\prime}\cdot C_{\ell}^{\nu_2, \prime}}}, 
\eeq
which are presented as red circles with black error bars in Figure~\ref{fig:xcorr}.  
Note that $C_{\ell}^{\prime}$ here represents ``CIB-only" bandpowers -- including only the CIB and any tSZ-CIB correlation. 
Thus the reported correlation coefficient will be sensitive to the magnitude of the tSZ-CIB correlation as well as the actual CIB correlations. 
Suppressing the other components is necessary because each component will have its own correlation structure. 
In particular, the CMB is extremely bright on large angular scales in the SPT bands and perfectly correlated across frequencies. 
Thus, including CMB power would tend to bring the correlation between the SPT bands to unity at low-$\ell$. 
To create these CIB-only bandpowers, we subtract the best-fit values (including calibration factors from that point in the chain) from  \citetalias{george15} for the power from the CMB, tSZ, kSZ, radio galaxies and galactic cirrus from the measured bandpowers. 

Fig.~\ref{fig:xcorr} shows the resulting correlation coefficients as a function of multipole.
The observed correlations are largely  independent of angular scale, although there is a suggestion of a slope in the terms involving 220\,GHz and the SPIRE bands. 
There is a visual hint of a slope in the  95\,GHz correlation coefficients as well, which is  
intriguingly as the slope matches the expected direction for an anti-correlation between the tSZ and CIB power (i.e.~the tSZ-CIB anti-correlation should reduce the cross-correlation coefficient as defined here around $\ell\sim3000$). 
However, we stress that the uncertainties are large, and the predicted variations in the \citetalias{george15} model on these scales is small compared to the current error bars. 
In the SPIRE-only maps, correlation coefficients are effectively independent of angular scale.
The measured correlation coefficients, averaged across $\ell \in[2000,7000]$, are reported in Table~\ref{tab:correl}. 
For the SPIRE bands, we find correlation values of $<\xi> =0.970\pm0.010$, $0.861\pm 0.012$, $0.9551\pm 0.0099$ for $600\times 857$, $600\times 1200$, and $857\times 1200$\,GHz, respectively.  
These values agree with those found by \citet{viero13a}. 
We report the average correlation between all frequency pairs in Table~\ref{tab:correl}.

\begin{table*}[]
\center
\caption{Mean correlation coefficients between each frequency pair }
 \begin{tabular}{c|cccccc}
 &95\,GHz & 150\,GHz &220\,GHz & 600\,GHz  & 857\,GHz  & 1200\,GHz \\
 \hline
  95\,GHz & 1  & $0.53\pm0.43$ &  $0.68\pm 0.55$ & $0.46\pm 0.36$ & $0.40\pm 0.32$ & $0.36 \pm 0.28$ \\
 150\,GHz & -  & 1 &   $0.976\pm 0.039$ & $0.879\pm0.025$ & $0.789\pm 0.022$ & $0.650\pm 0.020$ \\
  220\,GHz & -  & - &   1 & $0.879\pm 0.018$ & $0.786\pm 0.016$ & $0.643 \pm 0.014$ \\
    600\,GHz & -  & - &   - & 1 & $0.970\pm0.010$ & $0.861\pm 0.012$ \\
      857\,GHz & -  & - &   - & - & 1 & $0.9551\pm 0.0099$ \\
       1200\,GHz & -  & - &   - & - & - & 1 \\
        \end{tabular}
        \tablecomments{\label{tab:correl} Correlation coefficients from the weighted average of bandpowers on angular scales $\ell\in[2000,7000]$. 
        Except for 220\,GHz, the correlation is consistent with being independent of angular scale. 
        Unsurprisingly, the correlation decreases for frequency bands that are further apart. 
         }
\end{table*}

%%%%%%%%%%%%%%%%%%%%%%%%%%
\begin{figure}[h]
\centering
\includegraphics[width=0.46\textwidth]{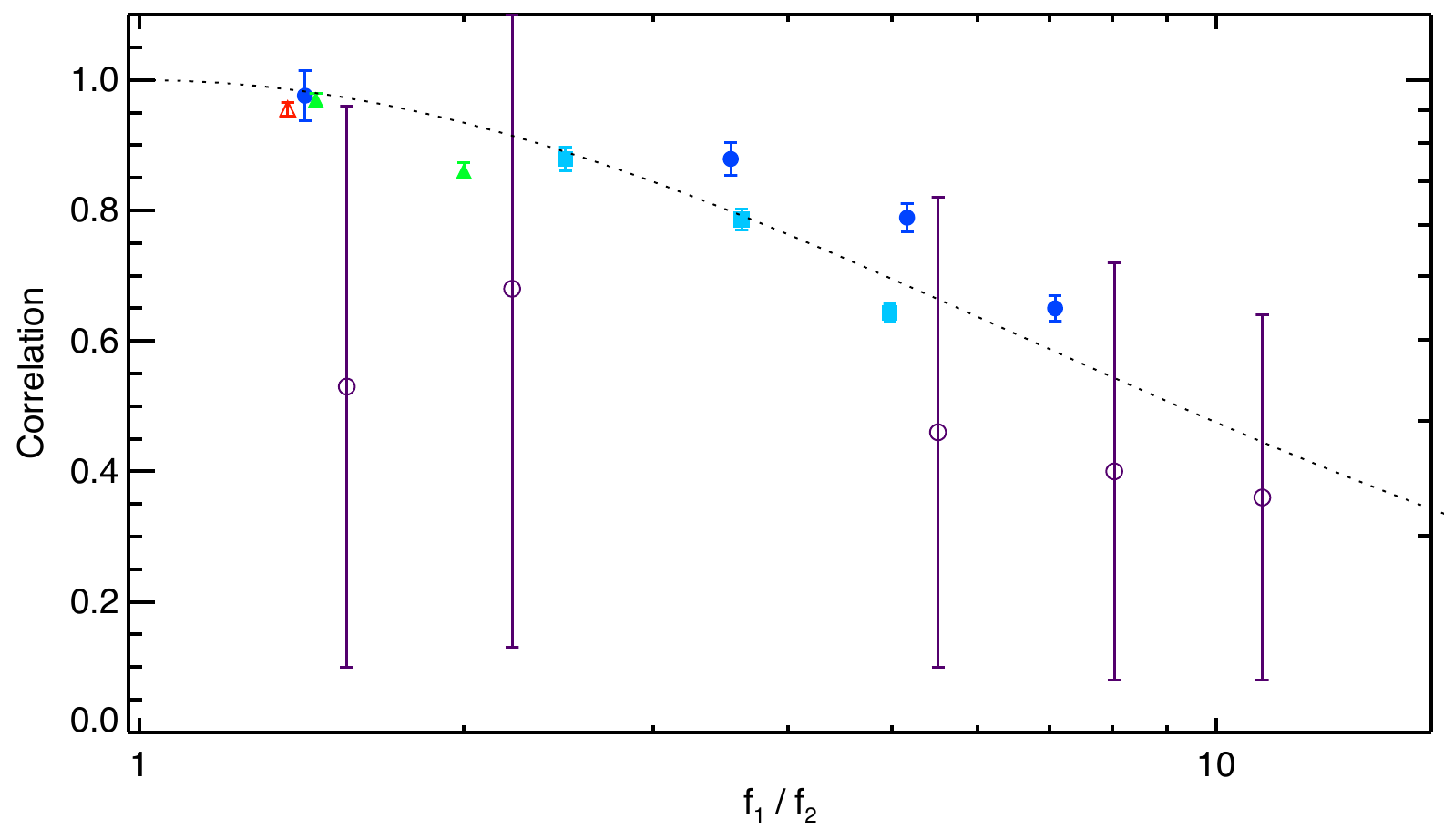}
\hspace{3mm}
\caption{
The measured correlation between bands declines quickly as the frequency separation increases. 
The plot shows the correlation coefficients, as defined in Eqn.~\ref{eqn:correl}, averaged across $\ell\in[2000,7000]$ as a function of ratio of the frequency for each band. 
The dotted line shows a toy model prediction for this correlation if the $\beta$ in the modified black-body for each CIB galaxy is drawn from a Normal distribution with $\sigma=0.53$. 
While the toy model is a poor fit to the measured correlation coefficients ($\chi^2=109$ for 14 d.o.f.), it does capture the qualitative behavior of the correlation coefficient. 
Frequency pairs where the lower frequency band is 95\,GHz are marked by open purple circles, 150\,GHz by filled blue circles, 220\,GHz by filled turquoise squares, 600\,GHz by filled green triangles, and 857\,GHz by a open red triangle. 
}
\label{fig:correff}
\end{figure}
%%%%%%%%%%%%%%%%%%%%%%%%%%

Our expectation is that the CIB should be highly correlated in neighboring frequency bands, with the correlation dropping as the frequency separation increases. 
Even if all DSFGs had identical spectral energy distributions (SEDs) in their local restframe, their observed SEDs would peak at different frequencies based on their individual redshifts. 
Effectively different frequency bands are weighted towards DSFGs at different redshifts. 
As the frequency separation increases, there is less overlap in which DSFGs are contributing to each band's power spectrum and more decorrelation. 
Variations in the restframe SED will also add to the decorrelation. 
We can measure this decoherence in the data, as shown in Fig.~\ref{fig:correff}. 
In this plot, the average (from $\ell=2000$ to 7000) cross-correlation coefficients are shown for each of the 15 unique frequency pairs. 
The colored symbol encodes the lower frequency in the pair, e.g., open purple circles for the 5 pairs involving the 95\,GHz band and an open red triangle for the last pair between 857\,GHz and 1200\,GHz. 
The x-axis encodes the logarithmic frequency difference between the two bands in the pair. 
An overly simplistic model for the decoherence is simply assuming each galaxy has a different $\beta$ for its modified black-body, drawn from a Normal distribution with some variance $\sigma^2$. 
The correlation factors are insensitive to $T$ and $\beta$, so we fix these parameters to $T=18\,{\rm K}$ and $\beta=2.0$. 
We conduct a fit to this toy model, and find  $\sigma^2 = 0.277 \pm 0.009$ (or $\sigma = 0.53$). 
The quoted error on $\sigma^2$ is based on the change that increases the $\chi^2$ by one; this is likely to be optimistic given the poor overall fit (the best-fit $\chi^2=109$). 
The dotted line in  Fig.~\ref{fig:correff} shows the predictions for the best-fit to this toy model. 
The data are qualitatively consistent with the degree of correlation depending only on the ratio of the frequencies, as would be expected for variations in the power law spectral index between DSFGs. 
We stress that one should be very cautious in interpreting the toy model physically -- the toy model does not include a DSFG redshift distribution and vacuums up the effects of different redshifts into variations in $\beta$.

We can compare the measured cross-correlation coefficients  to the predictions of the CIB modelling work in \citet{bethermin13}. 
Table 1 in that work presents predicted correlation coefficients for the clustering component at $\ell=1000$ and the Poisson component at higher multipoles; the measurement in this work across $\ell= 2000$ to 7000 should fall in between the two. 
There is also the caveat that the predictions are for frequencies close (but not identical) to the actual band centers.
The agreement is reasonable at a qualitative level (a careful statistical comparison is difficult given the caveats). 
We measure a correlation from 600 to 857 GHz (recall the effective CIB band centers are 546 and 796\,GHz) of $0.970\pm0.010$. \citet{bethermin13} predict the correlation between 545 and 857\,GHz to be 0.916 (Poisson) to 0.971 ($\ell=1000$). 
The prediction for 220 to 857\,GHz is  0.785 (Poisson) to 0.902 ($\ell=1000$); the actual observed correlation is $0.786\pm0.016$. 
The prediction for 100 to 857\,GHz is  0.743 (Poisson) to 0.919 ($\ell=1000$); the actual observed correlation is $0.40\pm0.32$. 
There is a suggestion, especially in the 95\,GHz band, that the observed correlations are lower than predicted, however the significance is low at the current level of uncertainty. 
A low value would not necessarily point to a flaw in the CIB modelling either, since the the model did not include the expected anti-correlation of the tSZ and CIB. 
The anti-correlation of the tSZ and CIB is most significant in the cross-frequency spectra including 95\,GHz. 
We conclude that the observed correlations are in rough agreement with the predicted correlations of the \citet{bethermin13} CIB model.

%#######################
\subsection{Comparison to Previous Measurements}
\label{sec:measurements}
%#######################

It is interesting to compare the spectra from this work with other CIB studies. 
Figure~\ref{fig:raw_ps} shows other recent measurements along with the $\rm SPIRE\times SPT$ bandpowers.  
CIB measurements from SPIRE \citep[blue diamonds from 600 to 1200\,GHz;]{viero13a} and \emph{Planck} \citep[green squares from 600 to 857\,GHz][]{planck13-30} are in excellent agreement.     
Note that we have scaled the \emph{Planck} bandpowers at 660 and 857\,GHz to account for differences in the passbands of SPIRE and \emph{Planck}.

For the $\rm CIB\times CMB$ measurements, we show measurements from $\rm BLAST\times ACT$ \citep[orange crosses over 150--$220\times600$--1200\,GHz;][]{hajian12}, and again from \emph{Planck} (green squares over 150--$220\times600$--857\,GHz).
The different datasets are in excellent agreement. 
As before, \emph{Planck} values are adjusted to account for passband differences. 
It is notable how the $\rm SPIRE\times SPT$ and \emph{Planck} spectra complement each other, with the two data sets overlapping at $\ell\sim 2000$.  
This is because the \emph{Planck} points are limited on smaller angular scales by the $\sim 5\, \rm arcmin$ beam, while the $\rm SPIRE\times SPT$ points are limited on larger scales by presence of CMB in the SPT maps -- which acts as a noise term and, unlike in the \emph{Planck} measurements, is not subtracted. 
The extra variance due to the CMB can be substantially larger than the expected CMB power in the $\rm SPIRE\times SPT$ cross-spectrum because the variance includes terms proportional to the $\rm SPT\times SPT$ auto-spectrum (which is dominated by the CMB). 
For instance, the variance of the  $1200\times 150$\,GHz cross-spectrum has terms proportional to, $C^{150\times 150}_\ell C^{1200\times 1200}_\ell + (C^{150\times 1200}_\ell)^2$, and the CMB is extremely bright in $C^{150\times 150}_\ell$ despite being negligible in the other terms.
The magnitude of this extra variance due to the CMB is  shown by a red dashed line.

% ======================
\section{Conclusion}
\label{sec:conclusion}
% ======================

We have presented high-angular-resolution temperature power spectra from a cross-frequency power spectrum analysis of an approximately 100 \sqdeg{} field observed with \herschelspire{} and the SPT. 
We show for the first time the cross-frequency spectra between the CIB (at 600, 857, and 1200\,GHz; or 500, 350, and 250\rmicron)
and maps at 95\,GHz (3.2\,mm). 
These cross-spectra show the qualitative behavior predicted for the expected anti-correlation between the CIB and tSZ signals.  
From $2000<\ell<11000$, the reported bandpowers also dramatically improve upon earlier measurements  of the cross-spectra between the CIB and maps near the peak of the CMB at 150 and 220\,GHz. 
The measurements at $\ell < 2000$ are consistent with earlier \planck{} results. 

We also measure the observed correlation between the CIB maps across this 12-fold frequency range (from 95 to 1200 GHz). 
Decorrelation between frequency bands is expected to be introduced by  differences in the SED from one DSFG to another. 
This model predicts a high degree of correlation for nearby bands, dropping towards zero with increasing frequency separation. 
The observed correlations are largely consistent with this picture.

The bandpowers, covariance matrix and bandpower window functions from this work are publicly available\footnote{http://pole.uchicago.edu/public/data/viero18/}.
% and http://lambda.gsfc.nasa.gov/product/spt/XXX.cfm}. 
The data presented here may be of interest to testing halo-occupation models for the CIB, for probing the correlation between galaxy clusters and the DSFGs, and for putting limits on the epoch of reionization through the kSZ effect. 

The SPT-3G survey began taking data in 2017. 
With roughly ten-fold more detectors than SPT-SZ and equal detector counts at 95, 150 and 220\,GHz, SPT-3G has embarked upon a 1500\,\sqdeg{} survey  (including this field) that will have significantly lower noise levels than the SPT-SZ survey. 
These data can be used to make further improvements on our modelling of the CIB and tSZ-CIB correlations.

\newpage
\begin{acknowledgments}
    	The South Pole Telescope program is supported by the National Science Foundation through grant PLR-1248097. Partial support is also provided by the NSF Physics Frontier Center grant PHY-0114422 to the Kavli Institute of Cosmological Physics at the University of Chicago, the Kavli Foundation, and the Gordon and Betty Moore Foundation through Grant GBMF\#947 to the University of Chicago. 
The McGill authors acknowledge funding from the Natural Sciences and Engineering Research Council of Canada, Canadian Institute for Advanced Research, and Canada Research Chairs program.
	G. S. acknowledges support from the Fonds de recherche du Qu\'ebec - Nature et technologies.
BB has been supported by the Fermi Research Alliance, LLC under Contract No. DE-AC02-07CH11359 with the U.S. Department of Energy, Office of Science, Office of High Energy Physics.
CR acknowledges support from Australian Research Council's Discovery Projects scheme (DP150103208).
This research used resources of the National Energy Research Scientific Computing Center, which is supported by the Office of Science of the U.S. Department of Energy under Contract No. DE-AC02-05CH11231. 
We acknowledge the use of the Legacy Archive for Microwave Background Data Analysis (LAMBDA). Support for LAMBDA is provided by the NASA Office of Space Science.

SPIRE has been developed by a consortium of institutes led
by Cardiff Univ. (UK) and including: Univ. Lethbridge (Canada);
NAOC (China); CEA, LAM (France); IFSI, Univ. Padua (Italy);
IAC (Spain); Stockholm Observatory (Sweden); Imperial College
London, RAL, UCL-MSSL, UKATC, Univ. Sussex (UK); and Caltech,
JPL, NHSC, Univ. Colorado (USA). This development has been
supported by national funding agencies: CSA (Canada); NAOC
(China); CEA, CNES, CNRS (France); ASI (Italy); MCINN (Spain);
SNSB (Sweden); STFC, UKSA (UK); and NASA (USA).
SPIRE maps were observed as part of the {\tt OT1\_jcarls01\_3} program, and made from the following OBSIDS: 1342232364-5, 1342245412-4, 1342245430-2, 1342245510-2, 1342245547.

Argonne National Laboratory work was supported under U.S. Department of Energy contract DE-AC02-06CH11357. 
\end{acknowledgments}

%APPENDIX

%----------------- TABLE BANDPOWERS ------------------------------
\begin{table*}
 \footnotesize
 \centering
\caption{Auto- and Cross- Frequency Power Spectra and $1 \sigma$ Uncertainties [$\rm Jy^2\, sr^{-1}$]}
 \begin{tabular}{lc|ccc| cc| c}
 \hline
 \hline
 $\ell_{\rm min} - \ell_{\rm max}$ & $\ell_{\rm eff}$ &  $C_\ell^{95}$ &  $C_\ell^{95\times150}$ &  $C_\ell^{95\times220}$ &  $C_\ell^{150}$ &  $C_\ell^{150\times220}$ & $C_\ell^{220}$ \\
\hline
 2001 -  2200 &  2068 & $ 18.33\pm0.93 $ & $ 30.08\pm1.29 $ & $ 34.33\pm1.81 $ & $ 52.08\pm2.16 $ & $ 65.16\pm3.03 $ & $ 94.55\pm5.67 $ \\
 2201 -  2500 &  2323 & $  9.14\pm0.48 $ & $ 13.84\pm0.52 $ & $ 15.64\pm0.86 $ & $ 23.86\pm0.82 $ & $ 31.22\pm1.25 $ & $ 51.74\pm2.98 $ \\
 2501 -  2800 &  2630 & $  4.04\pm0.31 $ & $  6.43\pm0.30 $ & $  8.29\pm0.61 $ & $ 11.84\pm0.45 $ & $ 17.79\pm0.73 $ & $ 34.95\pm1.90 $ \\
 2801 -  3100 &  2932 & $  2.12\pm0.31 $ & $  3.19\pm0.20 $ & $  3.83\pm0.42 $ & $  6.19\pm0.25 $ & $ 10.39\pm0.48 $ & $ 23.61\pm1.74 $ \\
 3101 -  3500 &  3288 & $  1.36\pm0.24 $ & $  1.90\pm0.13 $ & $  1.97\pm0.40 $ & $  3.59\pm0.15 $ & $  7.35\pm0.35 $ & $ 23.24\pm1.28 $ \\
 3501 -  3900 &  3690 & $  0.63\pm0.24 $ & $  1.07\pm0.11 $ & $  1.67\pm0.39 $ & $  2.50\pm0.12 $ & $  5.84\pm0.26 $ & $ 19.64\pm0.99 $ \\
 3901 -  4400 &  4143 & $  1.17\pm0.25 $ & $  0.73\pm0.11 $ & $  1.86\pm0.28 $ & $  1.88\pm0.09 $ & $  5.02\pm0.23 $ & $ 17.49\pm0.96 $ \\
 4401 -  4900 &  4645 & $  0.26\pm0.27 $ & $  0.65\pm0.10 $ & $  1.43\pm0.36 $ & $  1.67\pm0.09 $ & $  4.70\pm0.21 $ & $ 17.09\pm0.93 $ \\
 4901 -  5500 &  5198 & $  0.67\pm0.30 $ & $  0.64\pm0.10 $ & $  0.87\pm0.32 $ & $  1.48\pm0.07 $ & $  4.53\pm0.17 $ & $ 16.81\pm0.73 $ \\
 5501 -  6200 &  5851 & $  0.37\pm0.36 $ & $  0.40\pm0.10 $ & $  1.05\pm0.33 $ & $  1.40\pm0.06 $ & $  4.24\pm0.15 $ & $ 16.14\pm0.67 $ \\
 6201 -  7000 &  6604 & $  0.10\pm0.46 $ & $  0.51\pm0.12 $ & $  1.16\pm0.39 $ & $  1.38\pm0.07 $ & $  4.20\pm0.16 $ & $ 14.81\pm0.68 $ \\
 7001 -  7800 &  7406 & $  1.01\pm0.66 $ & $  0.43\pm0.16 $ & $  0.86\pm0.49 $ & $  1.33\pm0.08 $ & $  3.72\pm0.18 $ & $ 15.07\pm0.77 $ \\
 7801 -  8800 &  8309 & -  & $  0.32\pm0.20 $ & $  1.88\pm0.62 $ & $  1.09\pm0.09 $ & $  4.18\pm0.19 $ & $ 14.29\pm0.82 $ \\
 8801 -  9800 &  9312 & -  & $  0.44\pm0.32 $ & $  1.39\pm0.95 $ & $  1.31\pm0.12 $ & $  4.22\pm0.25 $ & $ 14.06\pm1.04 $ \\
 9801 - 11000 & 10416 & -  & -  & -  & $  1.23\pm0.15 $ & $  3.74\pm0.30 $ & $ 15.06\pm1.21 $ \\

 \end{tabular}
  \footnotesize 
\tablecomments{Bandpowers for the SPT frequency bands in units of $\rm Jy^2\, sr^{-1}$ under the photometric convention $\nu I_\nu = $ constant. The reported uncertainties are based on the diagonal of the covariance matrix, and do not include beam or calibration uncertainties. }
\label{tab:bandpowers}
\end{table*}
 
\begin{table*}
 \tiny
 \centering
\caption{Auto- and Cross- Frequency Power Spectra and $1 \sigma$ Uncertainties [$\rm Jy^2\, sr^{-1}$]}
 \begin{tabular}{lc | ccc | ccc | ccc}
 \hline
 \hline
 $\ell_{\rm min} - \ell_{\rm max}$ & $\ell_{\rm eff}$ &  $C_\ell^{95\times600}$ &  $C_\ell^{95\times857}$ &  $C_\ell^{95\times1200}$ &  $C_\ell^{150\times600}$ &  $C_\ell^{150\times857}$ &  $C_\ell^{150\times1200}$ &  $C_\ell^{220\times600}$ &  $C_\ell^{220\times857}$ &  $C_\ell^{220\times1200}$ \\
\hline
  801 -  1000 &   900 & -  & -  & -  & -  & -  & -  & $   913.\pm 472. $ & $  1281.\pm 597. $ & $  1276.\pm 757. $ \\
 1001 -  1200 &  1100 & -  & -  & -  & $   231.\pm 141. $ & $   116.\pm 192. $ & $   317.\pm 254. $ & $   742.\pm 191. $ & $   706.\pm 268. $ & $   914.\pm 362. $ \\
 1201 -  1400 &  1300 & -  & -  & -  & $   133.\pm  82. $ & $   182.\pm 121. $ & $   263.\pm 153. $ & $   491.\pm 115. $ & $   749.\pm 172. $ & $   937.\pm 217. $ \\
 1401 -  1600 &  1500 & -  & -  & -  & $    69.\pm  51. $ & $    44.\pm  72. $ & $    94.\pm 100. $ & $   360.\pm  73. $ & $   413.\pm 103. $ & $   450.\pm 145. $ \\
 1601 -  1800 &  1700 & -  & -  & -  & $    86.\pm  29. $ & $   109.\pm  43. $ & $   117.\pm  61. $ & $   469.\pm  46. $ & $   658.\pm  66. $ & $   719.\pm  94. $ \\
 1801 -  2000 &  1900 & -  & -  & -  & $   123.\pm  19. $ & $   171.\pm  27. $ & $   134.\pm  40. $ & $   385.\pm  31. $ & $   535.\pm  45. $ & $   495.\pm  63. $ \\
 2001 -  2200 &  2100 & $   13.3\pm 8.5 $ & $    24.\pm  12. $ & $    18.\pm  18. $ & $    90.\pm  14. $ & $   142.\pm  21. $ & $   127.\pm  29. $ & $   335.\pm  23. $ & $   496.\pm  34. $ & $   477.\pm  48. $ \\
 2201 -  2500 &  2350 & $    3.4\pm 4.7 $ & $    1.5\pm 6.9 $ & $     2.\pm  10. $ & $   65.9\pm 7.6 $ & $    80.\pm  11. $ & $    85.\pm  15. $ & $   300.\pm  16. $ & $   411.\pm  23. $ & $   427.\pm  31. $ \\
 2501 -  2800 &  2650 & $    8.3\pm 3.8 $ & $   15.9\pm 5.6 $ & $   13.8\pm 8.0 $ & $   76.7\pm 5.3 $ & $  110.4\pm 7.8 $ & $   113.\pm  10. $ & $   273.\pm  12. $ & $   400.\pm  17. $ & $   407.\pm  22. $ \\
 2801 -  3100 &  2950 & $   12.2\pm 3.0 $ & $   13.0\pm 4.8 $ & $   10.2\pm 6.8 $ & $   71.4\pm 3.7 $ & $   99.6\pm 5.5 $ & $   98.6\pm 7.7 $ & $  264.9\pm 9.3 $ & $   367.\pm  14. $ & $   377.\pm  19. $ \\
 3101 -  3500 &  3300 & $   10.6\pm 2.7 $ & $   11.6\pm 4.1 $ & $   14.2\pm 6.0 $ & $   65.3\pm 2.6 $ & $   89.8\pm 3.9 $ & $   91.8\pm 5.3 $ & $  244.9\pm 6.8 $ & $   335.\pm  10. $ & $   341.\pm  14. $ \\
 3501 -  3900 &  3700 & $   10.3\pm 2.3 $ & $   15.8\pm 3.6 $ & $   17.5\pm 5.3 $ & $   60.2\pm 2.1 $ & $   84.7\pm 3.2 $ & $   84.6\pm 4.7 $ & $  234.0\pm 6.2 $ & $  336.8\pm 8.8 $ & $   349.\pm  12. $ \\
 3901 -  4400 &  4150 & $    7.9\pm 2.1 $ & $    9.1\pm 3.1 $ & $    6.3\pm 4.7 $ & $   56.6\pm 1.9 $ & $   81.4\pm 2.7 $ & $   81.9\pm 3.8 $ & $  217.3\pm 5.2 $ & $  307.7\pm 8.0 $ & $   314.\pm  11. $ \\
 4401 -  4900 &  4650 & $    9.7\pm 2.1 $ & $   14.4\pm 3.1 $ & $   18.4\pm 5.0 $ & $   55.9\pm 1.6 $ & $   77.3\pm 2.4 $ & $   79.7\pm 3.4 $ & $  214.7\pm 4.6 $ & $  305.3\pm 7.1 $ & $  305.0\pm10.0 $ \\
 4901 -  5500 &  5200 & $   10.4\pm 2.0 $ & $   17.4\pm 3.1 $ & $   18.7\pm 4.7 $ & $   55.2\pm 1.4 $ & $   75.7\pm 2.0 $ & $   78.2\pm 3.0 $ & $  202.5\pm 3.9 $ & $  277.4\pm 5.7 $ & $  280.2\pm 8.5 $ \\
 5501 -  6200 &  5849 & $   11.1\pm 2.0 $ & $   15.3\pm 3.1 $ & $   15.2\pm 4.6 $ & $   52.2\pm 1.2 $ & $   72.6\pm 1.8 $ & $   75.1\pm 2.8 $ & $  198.3\pm 3.3 $ & $  273.2\pm 5.0 $ & $  281.4\pm 7.4 $ \\
 6201 -  7000 &  6599 & $    7.1\pm 2.2 $ & $   12.3\pm 3.3 $ & $   16.1\pm 4.9 $ & $   51.5\pm 1.2 $ & $   71.7\pm 1.8 $ & $   75.3\pm 2.6 $ & $  182.6\pm 3.2 $ & $  254.2\pm 4.9 $ & $  261.5\pm 7.1 $ \\
 7001 -  7800 &  7399 & $    6.6\pm 2.5 $ & $    9.6\pm 4.0 $ & $   15.1\pm 5.9 $ & $   49.9\pm 1.2 $ & $   71.2\pm 1.7 $ & $   73.7\pm 2.6 $ & $  184.8\pm 3.4 $ & $  248.5\pm 4.9 $ & $  247.4\pm 7.3 $ \\
 7801 -  8800 &  8299 & $    5.9\pm 2.7 $ & $    9.9\pm 4.1 $ & $    5.5\pm 6.2 $ & $   48.4\pm 1.1 $ & $   68.1\pm 1.6 $ & $   72.1\pm 2.4 $ & $  171.4\pm 3.0 $ & $  234.2\pm 4.3 $ & $  242.8\pm 6.4 $ \\
 8801 -  9800 &  9299 & $   20.1\pm 3.6 $ & $   20.9\pm 5.6 $ & $   24.7\pm 8.3 $ & $   46.4\pm 1.2 $ & $   64.6\pm 1.9 $ & $   67.5\pm 2.7 $ & $  170.8\pm 3.5 $ & $  232.4\pm 5.0 $ & $  235.4\pm 7.4 $ \\
 9801 - 11000 & 10399 & -  & -  & -  & $   47.9\pm 1.2 $ & $   65.5\pm 1.9 $ & $   68.9\pm 2.9 $ & $  169.0\pm 3.2 $ & $  232.7\pm 4.8 $ & $  238.3\pm 6.7 $ \\

 \end{tabular}
  \footnotesize 
\tablecomments{Bandpowers for the SPTxSPIRE frequency bands in units of $\rm Jy^2\, sr^{-1}$ under the photometric convention $\nu I_\nu = $ constant. The reported uncertainties are based on the diagonal of the covariance matrix, and do not include beam or calibration uncertainties. }
\label{tab:bandpowers2}
\end{table*}
 
\begin{table*}
 \footnotesize
 \centering
\caption{Auto- and Cross- Frequency Power Spectra and $1 \sigma$ Uncertainties [$\rm 1000~Jy^2\, sr^{-1}$]}
 \begin{tabular}{lc|ccc| cc| c}
 \hline
 \hline
 $\ell_{\rm min} - \ell_{\rm max}$ & $\ell_{\rm eff}$ &  $C_\ell^{600\times600}$ &  $C_\ell^{600\times857}$ &  $C_\ell^{600\times1200}$ &   $C_\ell^{857\times857}$ &  $C_\ell^{857\times1200}$ &   $C_\ell^{1200\times1200}$ \\
\hline
  601 -   800 &   700 & $   18.9\pm 2.4 $ & $   26.5\pm 3.7 $ & $   30.8\pm 5.3 $ & $   39.3\pm 5.4 $ & $   48.2\pm 7.7 $ & $    57.\pm  12. $ \\
  801 -  1000 &   900 & $   12.4\pm 1.5 $ & $   19.0\pm 2.1 $ & $   21.0\pm 3.0 $ & $   30.4\pm 2.8 $ & $   34.9\pm 3.9 $ & $   43.3\pm 5.6 $ \\
 1001 -  1200 &  1100 & $   8.88\pm 0.96 $ & $   13.5\pm 1.4 $ & $   14.0\pm 2.1 $ & $   22.8\pm 1.8 $ & $   25.0\pm 2.5 $ & $   31.9\pm 3.5 $ \\
 1201 -  1400 &  1300 & $   7.44\pm 0.66 $ & $  12.01\pm 0.92 $ & $   13.0\pm 1.4 $ & $   18.9\pm 1.2 $ & $   22.7\pm 1.7 $ & $   28.6\pm 2.2 $ \\
 1401 -  1600 &  1500 & $   7.03\pm 0.48 $ & $  10.52\pm 0.67 $ & $   10.9\pm 1.0 $ & $  17.45\pm 0.90 $ & $   19.6\pm 1.3 $ & $   24.1\pm 1.6 $ \\
 1601 -  1800 &  1700 & $   6.42\pm 0.37 $ & $   9.54\pm 0.57 $ & $  10.62\pm 0.86 $ & $  15.55\pm 0.74 $ & $   18.4\pm 1.0 $ & $   23.9\pm 1.3 $ \\
 1801 -  2000 &  1900 & $   6.12\pm 0.30 $ & $   9.11\pm 0.43 $ & $  10.03\pm 0.62 $ & $  14.31\pm 0.57 $ & $  16.55\pm 0.74 $ & $  21.05\pm 0.97 $ \\
 2001 -  2200 &  2100 & $   5.22\pm 0.23 $ & $   7.89\pm 0.35 $ & $   8.76\pm 0.54 $ & $  12.72\pm 0.46 $ & $  14.75\pm 0.63 $ & $  18.91\pm 0.81 $ \\
 2201 -  2500 &  2350 & $   4.67\pm 0.16 $ & $   6.96\pm 0.26 $ & $   7.80\pm 0.39 $ & $  11.20\pm 0.34 $ & $  13.29\pm 0.46 $ & $  17.28\pm 0.57 $ \\
 2501 -  2800 &  2650 & $   4.57\pm 0.13 $ & $   6.97\pm 0.20 $ & $   7.67\pm 0.30 $ & $  11.22\pm 0.25 $ & $  13.21\pm 0.35 $ & $  16.87\pm 0.44 $ \\
 2801 -  3100 &  2950 & $   4.27\pm 0.11 $ & $   6.35\pm 0.17 $ & $   6.98\pm 0.27 $ & $   9.91\pm 0.22 $ & $  11.69\pm 0.33 $ & $  14.99\pm 0.40 $ \\
 3101 -  3500 &  3300 & $  3.797\pm0.084 $ & $   5.79\pm 0.13 $ & $   6.32\pm 0.21 $ & $   9.32\pm 0.17 $ & $  10.94\pm 0.25 $ & $  14.04\pm 0.31 $ \\
 3501 -  3900 &  3700 & $  3.717\pm0.070 $ & $   5.57\pm 0.12 $ & $   6.14\pm 0.18 $ & $   8.88\pm 0.14 $ & $  10.59\pm 0.22 $ & $  13.77\pm 0.26 $ \\
 3901 -  4400 &  4150 & $  3.509\pm0.055 $ & $  5.375\pm0.099 $ & $   5.91\pm 0.16 $ & $   8.72\pm 0.13 $ & $  10.29\pm 0.19 $ & $  13.31\pm 0.23 $ \\
 4401 -  4900 &  4650 & $  3.428\pm0.052 $ & $  5.175\pm0.087 $ & $   5.72\pm 0.15 $ & $   8.30\pm 0.11 $ & $   9.86\pm 0.18 $ & $  12.80\pm 0.22 $ \\
 4901 -  5500 &  5200 & $  3.267\pm0.044 $ & $  4.898\pm0.081 $ & $   5.42\pm 0.13 $ & $   7.82\pm 0.10 $ & $   9.33\pm 0.16 $ & $  12.18\pm 0.19 $ \\
 5501 -  6200 &  5849 & $  3.103\pm0.039 $ & $  4.681\pm0.074 $ & $   5.19\pm 0.13 $ & $  7.460\pm0.093 $ & $   8.90\pm 0.16 $ & $  11.68\pm 0.19 $ \\
 6201 -  7000 &  6599 & $  3.030\pm0.035 $ & $  4.576\pm0.061 $ & $   5.09\pm 0.11 $ & $  7.350\pm0.076 $ & $   8.82\pm 0.13 $ & $  11.59\pm 0.15 $ \\
 7001 -  7800 &  7399 & $  2.963\pm0.033 $ & $  4.485\pm0.060 $ & $   4.97\pm 0.11 $ & $  7.290\pm0.075 $ & $   8.66\pm 0.13 $ & $  11.36\pm 0.16 $ \\
 7801 -  8800 &  8299 & $  2.836\pm0.030 $ & $  4.274\pm0.057 $ & $   4.74\pm 0.11 $ & $  6.914\pm0.068 $ & $   8.28\pm 0.12 $ & $  10.91\pm 0.14 $ \\
 8801 -  9800 &  9299 & $  2.796\pm0.028 $ & $  4.182\pm0.056 $ & $   4.64\pm 0.10 $ & $  6.650\pm0.066 $ & $   8.00\pm 0.12 $ & $  10.56\pm 0.14 $ \\
 9801 - 11000 & 10399 & $  2.690\pm0.027 $ & $  4.079\pm0.054 $ & $   4.51\pm 0.11 $ & $  6.583\pm0.063 $ & $   7.89\pm 0.11 $ & $  10.43\pm 0.14 $ \\

 \end{tabular}
  \footnotesize 
\tablecomments{Bandpowers for the SPIRE frequency bands in units of ($\rm 1000~Jy^2\, sr^{-1}$) under the photometric convention $\nu I_\nu = $ constant. The reported uncertainties are based on the diagonal of the covariance matrix, and do not include beam or calibration uncertainties. }
\label{tab:bandpowers3}
\end{table*}

\appendix

\section{Signals in the maps}
\label{app:signals}

Anisotropy in the cosmic microwave and infrared backgrounds can be introduced by primordial fluctuations, emission from DSFGs and radio galaxies,  the kinematic and thermal SZ effects, and contamination from Galactic foregrounds. 
The power spectrum at a given wavelength  can be written as: 
\begin{equation}
\begin{array}{r@{}l}
    C_{\ell}^{\rm Tot}  &{}=  C_{\ell}^{\rm CMB} + C_{\ell}^{\rm CIB} + C_{\ell}^{\rm radio} + C_{\ell}^{\rm tSZ-CIB} \\
    &{} + C_{\ell}^{\rm tSZ} + C_{\ell}^{\rm kSZ} + C_{\ell}^{\rm cirrus}  
\end{array}
\label{eqn:total}
\end{equation}
where $C_{\ell}$ is the angular power spectrum at multipole $\ell$.\footnote{In some studies the angular power spectrum is expressed as a function of the angular wavenumber, $k$, which is related to multipole number $\ell$ as $\ell = 2\pi k$. }

Relative to the CMB, the CIB and Galactic cirrus power rises towards higher frequencies, while the radio galaxy and tSZ power is more significant at lower frequencies. 
Wavelengths longer than roughly 500\rmicron\ are more sensitive to higher-redshift DSFGs \citep[for a review see][]{casey14}.  
Cross-spectra between frequency bands provide additional information about how the signals are correlated between bands, and allow one to disentangle signals that might otherwise be degenerate with one another. 
As a practical matter, only the CIB (from DSFGs) and cirrus terms are significant at the SPIRE frequency bands.

%#######################
%\subsection{Galactic Cirrus}
\subsection{Galactic Foreground Terms}
\label{sec:cirrus_ps}
%#######################
Diffuse Galactic cirrus introduces a foreground signal on large scales to the extragalactic submm/mm-wave power spectrum.  
Other Galactic foregrounds such as synchrotron and free-free emission are negligible in the area of the sky and frequencies considered here and thus are ignored.    

Galactic cirrus can be approximated as a power law 
\begin{equation}
C_{\ell}^{\rm cirrus}=P_{0,\lambda}\left( \frac{\ell}{\ell_0} \right)^{\alpha_{\rm c}},
\label{eqn:cirrus}
\end{equation}
with the amplitude $P_{0,\lambda}$ dependent on the observing frequency $\lambda$,
and the index $\alpha_{\rm c}$ independent of observing frequency \citep[e.g.,][]{gautier92,miv07}.   
In the first studies of high Galactic latitude cirrus, the index had been found to have a value of around $ -3.0$ \citep[e.g.,][]{gautier92}; however, more recent measurement have found that both the amplitude and index are quite direction-dependent, with values for $\alpha_{\rm c}$ ranging from 
$-1$ to $-4$ \citep[e.g.,][]{roy10, boothroyd11, bracco11}.

%Lastly, diffuse 
We also need to consider the spectral dependence of the Galactic cirrus. 
At submillimeter/millimeter wavelengths, cirrus emission follows a modified blackbody, $\nu^{\beta_c} B_{T_{\rm eff,c}}(\nu)$.
Here $B_{T_{\rm eff,c}}(\nu)$ is the Planck function for temperature $T_{\rm eff,c}$ and $\beta_c$ is the emissivity index \citep{draine84}.
Past studies have found $T_{\rm eff,c}=16-20\, \rm K$ and $\beta_c \approx 1.8$ \citep[e.g.,][]{bracco11}.

%#######################
\subsection{Cosmic infrared background (CIB)}
\label{sec:extragalactic}
%#######################

The CIB power term in Eqn.~\ref{eqn:total} is sourced by DSFGs and can be further split into Poisson and clustering components.  
The Poisson (or shot noise) term results from the discrete sampling of  dusty galaxies. 
The power spectrum of the Poisson term is independent of angular scale, with amplitude
\be
\label{eq:poisson}
 C_{\ell}^{\rm Poisson}  = \int^{S_{\rm cut}}_{0} S^2_{\nu} \frac{dN}{dS_{\nu}}\left(  S_{\nu} \right) d S_{\nu}, 
\ee 
where $dN/dS_{\nu}$ is the distribution of flux densities of sources, and $S_{\rm cut}$ is the level which sources with greater flux densities are masked.

The clustering term arises from overdensities in the background which trace the dark matter distribution. It can be described with a \lq halo model\rq\ approximation \citep[e.g.,][]{seljak00}, which consists of a linear (or 2-halo) term on large scales ($\ell \lsim 3500$ for the frequency bands in this work) and non-linear (or 1-halo) term on small scales ($\ell \gsim 3500$).  
The linear term is simply the dark matter power spectrum multiplied by an effective bias \citep[e.g.,][]{kaiser84}. 
The non-linear term is more subtle, as the bias can be both scale- and mass-dependent largely because %in dense environments 
galaxy properties (stellar masses, star-formation rates, bolometric luminosities, etc.) are tied to those of their host dark matter halos (halo mass, concentration, etc., e.g., \citealt{shang12,viero13a}).

\subsection{Radio Galaxies}
\label{sec:rg}

Radio galaxies also contribute to the Poisson term detailed in Eqn.~\ref{eq:poisson}. 
After masking sources above 6.4\,mJy at 150 GHz (see \S\ref{subsec:mask}), dusty galaxies are the dominant contributors to the Poisson term in all but the 95\,GHz band. 
At 95\,GHz, 80-90\% of the power is expected to be due to radio galaxies. 
As argued by \citetalias{george15} and others, we do not expect a detectable clustering term for the radio galaxies.

%#######################
\subsection{The thermal Sunyaev-Zel'dovich (tSZ) effect and tSZ-CIB correlations}
\label{sec:tsz}
%#######################

The galaxy clusters that give rise to the tSZ power spectrum similarly trace the dark matter distribution. 
Accurately modeling the tSZ power spectrum is challenging because of the complicated astrophysics affecting the intracluster medium in galaxy clusters; for instance non-thermal pressure support due to merger shocks. 
Different prescriptions for the astrophysics can change the predicted power for a given cosmology by up to 50\% \citep[e.g.,][]{shaw10,trac11,battaglia10,battaglia12,efstathiou12}. 

Additionally, since both galaxies and clusters trace the same dark matter, correlations between the two populations 
reveals insights into their relationship.  
The first to demonstrate the power of this relationship was \citet{addison12c}, who showed that the strength and shape of the cross-correlation spectrum --- which, again, appears as a negative signal --- is largely influenced by the HOD of star-forming galaxies in the most massive dark matter halos.  Furthermore, they show that since shorter wavelengths favor lower redshifts, that the strength of the effective anti-correlation is highest between 250\rmicron\ and CMB channels.   
The bandpowers reported in this work can be used to improve measurements of the tSZ-CIB correlation. 

%#######################
\subsection{The kinematic Sunyaev-Zel'dovich (kSZ) effect}
\label{sec:ksz}

The bulk motion of electrons can impart a Doppler shift upon scattered photons, an effect known as the kSZ effect \citep{sunyaev70b,sunyaev80}. 
The amplitude of the kSZ signal from a given volume scales as $(v/c) \delta_e$ where $c$ is the speed of light and $\delta_e$ is the over-density of free electrons. 
As a result, significant kSZ power is sourced during the epoch of reionization (when velocities and mass overdensities are small, but there is a high contrast in the free electron density between ionized and neutral regions) and at late times ($z < 2$, when velocities and mass overdensities have grown larger). 
We expect current models for the post-reionization kSZ power spectrum to be more accurate than tSZ models because, unlike the tSZ, the kSZ signal is not weighted by gas temperature. 
Thus, the kSZ spectrum is less dependent on the non-linear physics in dense halos. 
However, the kSZ power from reionization is highly uncertain, and offers a unique window into how reionization occurred \citep{mortonson10a, zahn12,mesinger12,george15}.

%#######################

\subsection{Cosmic microwave background (CMB)}
\label{sec:cmb}

Finally, the CMB is present  in all these maps.  
At SPIRE frequencies, the CMB is much fainter than the CIB and largely neglectable, but below 280\,GHz the CMB quickly overwhelms the CIB at large angular scales.  
 Although the two backgrounds are uncorrelated \citep[lensing notwithstanding, e.g.,][]{holder13}, 
 CMB sample variance introduces a large noise signal to the cross-correlations between SPT and SPIRE frequency bands, and washes out the CIB cross-correlation signal \citep[e.g.,][Appendix B]{hajian12}.  
 The \citet{planck13-30} overcame this limitation by subtracting the CMB directly from their maps, which they did by using their 100\,GHz map as a template and scaling the signal as a perfect 2.73\,K blackbody.  
 The drawback of this technique is that cross-frequency correlation measurements with 100\,GHz cannot be made --- a measurement that arguably contains the most novel information for the tSZ-CIB correlation.

\bibliographystyle{apj}
\bibliography{../../BIBTEX/spt}

% ======================
% ======================

\end{document}